\documentclass{article}
\usepackage{graphicx}
\usepackage{amsfonts,amssymb,amsbsy,amsmath}
\usepackage{amsfonts}
\usepackage{multirow} 
\usepackage{color}
\usepackage[utf8]{inputenc}

\newtheorem{thm}{Theorem}[section]

\begin{document}

\centerline{\Large\bf Change detection with adaptive sampling for binary responses}

\vspace{0.1in}

\noindent Yanqing YI$^1$\footnote{ Author to whom correspondence may be addressed. \\ {\it E-mail: yyi@mun.ca}} and Su-Fen YANG$^2$ 

\noindent{\it$^1$\it Faculty of Medicine, Memorial University of Newfoundland, St. John's, NL, Canada }

\noindent{\it$^2$\it College of Commerce, National Chengchi University, Taipei, Taiwan}

\noindent {\bf Abstract}

We propose using an adaptive sampling method to detect changes for a system with multiple lines. The adaptive sampling utilizes the information in responses to learn on which line is more likely to have a change thus allocating more units to the line. The learning process is formatted as a Markov decision process by integrating sampling information with likelihood ratio for changes to define rewards and the optimal sampling is approximated  by using the Bellman operator iteratively based on the average reward criterion. We demonstrate the performance of the proposed method for binary responses using the exact distribution method for adaptive sampling. Our numeric results show that the adaptive sampling samples more often the line that has a change and the statistical power to detect a change is better than those with the equal randomization for sample sizes of 20 or higher. When sample sizes increase or the difference between out-of-control and in-control probabilities increases, the adaptive sampling allocates higher proportion of units averagely to the line with a change and the statistical power to detect a change increases. 

\bigskip



\section{Introduction}

One of the sequential change detection problems is for a system with multiple lines of production where the sample units are limited due to high costs or availability of the resources. An example is the use of PCR test to screen COVID-19 infection before rapid tests were invented. In a bubble of a protection zone during the COVID-19 pandemic, travellers were sampled to test for COVID infection at airports or harbours with limited PCR boxes and health workers were screened to prevent the spread of the disease in health care facilities. With limited PCR boxes or resources for testing, the question is which hospital to sample and how often to test the sampled workers. We study the use of the information collected to adaptively sample lines that are more likely to have changes thus more observations are obtained for such lines to quickly detect a change in the system.

Adaptive sampling  has been utilized to survey species in geographical regions or ocean (Thompson and Seber 1996) while adaptive randomization has been used to allocate patients in clinical trials. Response adaptive randomization utilizes the accumulated information on treatment effect collected during a trial to sequentially change the randomization probability to allocate more patients to the potentially better treatment (Robertson et al. 2023). Adaptive sampling is employed to survey species  in ocean to estimate the population of the species . Such estimate may be used to manage issuing fishing licenses and to determine the quote of a license to preserve the sustainability of  natural resources. 
Recently, adaptive sampling/randomization was utilized for change point detection to minimize the regret function for multiple lines. Cao et al. (2019) considered a piecewise-stationary multi-arm bandit problem and proposed combining the uniform sampling  with the upper confidence bound exploration  to learn on the optimal arm for the mean regrets. Song and Fellouris (2024) proposed a framework for the optimization problem of minimizing the mean total cost regarding sequentially randomizing  strategy and optimal stopping, where the dynamic programming was used to solve the optimization problem and to establish the property of optimal stopping strategy based on the odds of a change occurring. A blocking of treatment assignment/sampling was used with the likelihood ratio statistic in Song and Fellouris (2024). Xu and Mei (2023) proposed the use of cyclic sampling in combining with the restricted parameter estimates in the CUSUM likelihood ratio statistic  to minimize the worst-case detection delay. The sampling strategies in those research are limited to uniform sampling, switching between blocks, or cyclic sampling. We consider a more general adaptive sampling strategy for change detection in this paper.

We formulate the sampling process for multiple lines as a Markov decision process using an average reward criterion under which the line with a potential change will be sampled sequentially with high sampling probability. The Markov decision process for adaptive sampling uses the information in observations to adaptively modify the sampling probability thus providing potential to allocate more units to the line with a change occurred. Considering small sample sizes, we demonstrate the performance of the sampling procedure for binary responses using the exact distribution of the process. For continuous responses, the exact distribution method is applicable after discretizing the state space (Fu et al. 2003). The Markov chain method has been used  for change detection with a single line (Page 1955, Champ and Woodall 1987, Fu et al. 2002). Although this method was also utilized to establish the upper and lower bounds for the procedure for change detection for multiple lines,  no adaptive sampling was considered (Nelson and Stepheson 1996, and Kuralmani et al. 2002). Another relative method to improve change detection is using variable sample sizes (VSS) with a larger or smaller size in the next stage, depending on the information on whether there is a change. Yang and Su (2006) developed VSS control charts to monitor two dependent processes and the method was extended to detect changes in cascade processes in Yang (2010).  We apply the Markov decision model proposed by Yi and Wang (2025) to study the optimal adaptive sampling to detect changes for multiple lines.

The remaining of this paper  is organized as follows. Section 2 introduces the adaptive sampling method built on the Markov decision model to detect changes for a system with multiple lines. Section 3 describes the Markov decision process for optimal sampling  and the exact distribution method for adaptive sampling with binary responses. Section 4 presents an algorithm for the optimal adaptive sampling to detect changes for binary responses and reports the numeric results on  the performance of the adaptive sampling. Section 5 concludes. 

\section{The adaptive sampling method}

In this section, we introduce the adaptive sampling method for change detection in a system with multiple lines. This method is based on a Markov decision model proposed by Yi and Wang (2025).

Suppose there are $k$ operating lines in a system. Let $\delta_{ij}$ be the indicator that line $j$ is sampled for unit/time $i, 1\leq j\leq k$. If line $j$ is sampled at time i, then an observation $y_{ij} \in \mathbb{S}$ is obtained. Let ${\cal S}$ be the $\sigma$-algebra on $\mathbb{S}$. Assume that $Y_{ij} \sim f_j(y) $ before a change and $Y_{ij} \sim g_j(y) $  after a change, where $f_j(y)$ and $g_j(y)$ are probability density/mass functions on $(S, {\cal S})$. Suppose at each time $i$, only one of the k lines is sampled. Denote $\textbf{y}_i=(\delta_{i1}y_{i1}, \delta_{i2}y_{i2}, \cdots, \delta_{ik}y_{ik})$ as the outcome of the sampled line at time $i$ and $\mbox{\boldmath$\delta$}_i=(\delta_{i1}, \delta_{i2}, \cdots, \delta_{ik})$, where $\delta_{ij}=1\; or \; 0$ and $\sum_{j=1}^k \delta_{ij}=1$. At time $i+1$, a sampling decision is made based on the cumulated information in the data $ h_{i}=(\mbox{\boldmath$\delta$}_{1}, \textbf{y}_{1}, \cdots, \mbox{\boldmath$\delta$}_{i}, \textbf{y}_{i} ).$ The  sampling probability is 
a sequence measurable maps $\mbox{\boldmath$\pi$} = \{\mbox{\boldmath$\pi$}_i,\; i =1,2,\cdots\}$ 
from $(\mathbb{H}_{i}, {\cal F}_{i})$ to $(\mathbb{A}, {\cal A}), i\geq 2$ such that for any
$\mbox{\boldmath$\delta$}_i\in \mathbb{A}^*,
\mathbb{A}^*=\{(\delta_{1}, \delta_{2}, \cdots, \delta_{k}): \sum_{j=1}^k\delta_{j}=1\},$ and $\mbox{\boldmath$\pi$}_i(\mathbb{A}^*|h_{i})=1 (\;\mbox{i.e.\;} \sum_{j=1}^k \pi_{ij} = 1),$
where   $H_1= \mathbb{A}^*\times \mathbb{S}, H_i=\{(h, \mbox{\boldmath$\delta$}, s), h\in \mathbb{H}_{i-1},
   \mbox{\boldmath$\delta$}\in \mathbb{A}^*, s\in\mathbb{S}\}, i\geq 2,$
  ${\cal F}_{i}$ is the $\sigma$ algebra on $H_i$, and $\mbox{\boldmath$\pi$}_1$ is pre-defined from the previous knowledge. 
  
Let $\mathbb{H}=\mathbb{A}^*\times\mathbb{S}\times \cdots,$ and ${\cal H}$
be the corresponding product $\sigma$-algebra. According to C. Ionescu Tulcea's theorem, for a sampling strategy $\mbox{\boldmath{$\pi$}},$
there is an unique probability measure $P_{\mbox{\boldmath$\pi$}}$ 
such that
\begin{eqnarray*}\label{bd}
& &  P_{\mbox{\boldmath$\pi$}}(\delta_i|h_{i-1})=\mbox{\boldmath$\pi$}_i(\delta_i|h_{i-1})\\
& &  P_{\mbox{\boldmath$\pi$}}(B|h_{i-1}, \delta_i)=\left\{
\begin{array}{ll}\prod_{j=1}^k\left(\int_{B_j} f_j(y)dy\right)^{\delta_{ij}} & \mbox{before a change}\\
 \prod_{j=1}^k\left(\int_{B_j} g_j(y)dy\right)^{\delta_{ij}} & \mbox{after a change}
 \end{array}\right.
\end{eqnarray*}
where 
$h_{i-1}\in \mathbb{H}_{(i-1)}, \mbox{\boldmath$\delta$}_{i}\in \mathbb{A}^*, B\in {\cal S},$ and $B_j$ is the
projection of $B$ on the $j^{th}$ component.

Let the reward function 
$r(\textbf{y}_i,\mbox{\boldmath{$\delta$}}_i) = \sum_{j=1}^k \delta_{ij} log\frac{g_j(y_{ij})}{f_j(y_{ij})} $ for unit $i$ with the sampling results $\mbox{\boldmath{$\delta$}}_i$ and response outcome $\textbf{y}_i.$ This reward function is related to the likelihood ratio test statistic. The sequential probability ratio test was proven to minimize the expected sample size on stopping to favour a simple alternative hypothesis among the tests with the same or smaller error probability of falsely rejecting the null hypothesis (Wald 1947, Chow et al. 1971). Built on the sequential testing theory, we use the rewards collected  to sequentially learn on which line is more likely to have a change happened thus determining the adaptive sampling probabilities. We consider the sampling strategy  $\Pi = \{\mbox{\boldmath$\pi$}=(\mbox{\boldmath$\pi$}_i): \mbox{\boldmath$\pi$}_i=(\pi_{i1}, \pi_{i2, }, \cdots, \pi_{ik}), \gamma \leq \pi_{ij} \leq 1-\gamma, j=1, 2, \cdots, k\}$, where $\gamma \leq 1/k$ is a pre-specified value. This type of sampling strategies provides a minimum of sampling probability $\gamma$ to explore for the possible changes among the $k$ lines. We use the average criterion $max_{\mbox{\boldmath$\pi$}\in \mbox{\boldmath$\Pi$}}\left [\liminf_{n \to \infty}\frac{E_{\mbox{\boldmath$\pi$}}\left (\sum_{i=1}^n r(\textbf{y}_i, \mbox{\boldmath{$\delta$}}_i)\right )}{n}\right ]$ to obtain the optimal sampling for change detection, where $E_{\mbox{\boldmath$\pi$}}$ is the expectation regarding the probability measure $P_{\mbox{\boldmath{$\pi$}}} .$  The sequential learning based on the average criterion was shown to have good small sample property to learn on the optimal sampling (Yi and Wang 2024). For a sampling strategy $\mbox{\boldmath$\pi$},$ the average reward  is $E_{\mbox{\boldmath$\pi$}}\left(\sum_{j=1}^k \frac{N_j}{n} r(j)\right)$, where $r(j)=E(r(Y, \mbox{\boldmath{$\delta$}}_{i}) | \delta_{ij}=1)$ and $N_j$ is the number of units sampled from line  $j$ when a total of $n$ units are sampled.  Denote $J^*=\{j^*: j^* =\arg max_{j}\{r(j)\}\}$. Let $|J^*|$ be the number of  lines in $J^*$. Applying the results in Yi and Wang (2024), we have the following.

\begin{thm}\label{op}

\begin{description}
\item (1) The optimal value of the average reward is $V^* =(1- (k-|J^*|)\gamma) \max_ j \{r(j)\} + \gamma \sum_{j\notin J^*} r(j)$.
\item (2) The optimal sampling is $\mbox{\boldmath$\pi$}^*=\{(\pi_{i,j}): \pi_{ij^*}=[1-(k-|J^*|)\gamma]/|J^*|, j^* \in J^* \}$ and $\pi_{ij}=\gamma, j\notin J^*.$
\end{description}
\end{thm}

The optimal sampling $\mbox{\boldmath$\pi$}^*$ depends on mean reward $r(j), j=1, \cdots,k.$ Due to the unknown parameter values in $r(j), j=1,2,\cdots, k,$ the information in the data $(\mbox{\boldmath$\delta$}_{1}, \textbf{y}_{1}, \cdots, \mbox{\boldmath$\delta$}_{i}, \textbf{y}_{i} )$ are used to learn on which line has a change  to determine the sampling probability for the next unit. This process can be formatted as a Markov decision process.

\section{The Markov decision process for adaptive sampling}

We introduce the Markov decision process for adaptive sampling to detect a change among the $k$ lines in this section. The process sequentially approximates the optimal sampling found under the average reward criterion. 


When $n$ units are sampled and the outcome responses are collected, the total rewards is
$$\sum_{i=1}^{n} r(y_i,\delta_i)=\sum_{i=1}^{n} \sum_{j=1}^k \delta_{ij} log\frac{g_j(y_{ij})}{f_j(y_{ij})}  =\sum_{j=1}^k W_j(n),$$

\noindent where $W_j(n)= \sum_{i=1}^n \delta_{ij} log\frac{g_j(y_{ij})}{f_j(y_{ij})}.$ After re-arranging index $i$, $W_j(n)=\sum_{\tilde{i}=1}^{N_j (n)}log\frac{g_j(y_{\tilde{i}j})}{f_j(y_{\tilde{i}j})} , j=1, 2, \cdots, k,$  where $N_j(n)=\sum_{i=1}^n \delta_{ij}$ is the number of units sampled from line $j$ and $\sum_{j=1}^k N_j(n)=n.$ Assume that the responses from line $j, j=1, 2, \cdots, k,$ are independent and follow the distribution $f_j(y)$ before a change point and $g_j(y)$ after the change. $W_j(n)$ is the logarithm of the likelihood ratio test statistic for line $j.$ We use the information in $W_j(n)$  to learn on which line is more likely to change for sampling decision for the next unit. Define $(W_j(n), j=1,2,\cdots,k)$ as the state and the process moves to state $(W_j(n+1))$
after the sampling and response $(\mbox{\boldmath$\delta$}_{n+1}, \textbf{y}_{n+1})$ for unit $n+1$ is obtained, where 
$W_j(n+1)=W_j(n)+r(\mbox{\boldmath$\delta$}_{n+1}, \textbf{y}_{n+1}), j=1,2,\cdots, k.$

The optimal sampling strategy is the one that maximizes the average reward. It can be approximated by using the Bellman operator $T u(w)$ under the average reward criterion,
$$T u(w) = \max_{\gamma \leq \pi \leq 1-\gamma } \sum_{j=1}^k\left\{ \pi(\delta_{.j}=1 |w)\left [r(j) + \int u(w+s) q(ds | w, \delta_{.j}=1)\right ] \right\} ,$$

\noindent where $\delta_{.j} $ is the sampling decision for line $j$ at any time $i$ and $q(s | w, \delta_{.j}=1)= g_j(s)$ after a change and $q(s | w, \delta_{.j}=1)=f_j(s)$ before a change. Since it is unknown whether a change occurs, we use a empirical density 
$\tilde{q}$ and estimate of $r(j)$ to iteratively obtain the optimal sequences $u_n(x),$ where 
$$
u_n(x) = \max_{\gamma \leq \pi \leq 1-\gamma }\sum_{j=1}^k\left\{ \pi'_n (\delta_{nj}=1 |x)\left [\hat{r}(j) + \int u_{n-1}(s) \tilde{q}(ds | x, j)\right ] \right\}.
$$

 This sampling process is a Markov decision process because it depends on $W_j(n)$, not on the previous states. 
The Bellman operator $T u(w)$ have been proven to be a contractor in that 
$sp(T u_1 - T u_2) (x) \leq (1-2\gamma) sp(u_1-u_2), $ for functions $u_1(x)$ and $u_2(x),$
\noindent where $sp(u)$ is the span semi-norm of $u(x)$, defined by $sp(u)=sup_x u(x) - inf_x u(x)$, and the policy identified using $u_n(x)$ converges to the optimal sampling $\mbox{\boldmath$\pi$}^*$ (Yi and Wang 2025). To simplify the notation, suppose that  $|J^*|=1$ in the following. To speed up the convergence to the optimal sampling, we use $\hat{u}_{n}=n\{(1-(k-1)\gamma) r(\theta_{\tilde{j}_n})+\gamma \sum_{j\ne \tilde{j}_n} r(\theta_j)\}$ to approximate $u_n$ in the Bellman operator $T,$ where $\tilde{j}_n$ is the best line with a potential change identified at iteration $n.$

The Markov decision process for adaptive sampling based on the Bellman operator $T$ works for continuous and binary outcomes. For binary outcomes, this process can be further simplified. Considering small to moderate sample sizes, we illustrate the Markov decision process for binary responses in the following. A similar algorithm can be established by discretizing the state space  for continuous responses (Fu et al. 2003, Lucas and Saccucci 1990). For binary outcomes, $W_j(n)= 
\sum_{\tilde{i}=1}^{N_j } log\frac{\theta_{j1}^{y_{\tilde{i}}}(1-\theta_{j1})^{1-y_{\tilde{i}}}}{\theta_{j0}^{y_{\tilde{i}}}(1-\theta_{j0})^{1-y_{\tilde{i}}}}
= S_j(n) log\bigg(\frac{\theta_{j1}/(1-\theta_{j1})}{\theta_{j0}/(1-\theta_{j0})}\bigg)+N_j(n) log\bigg(\frac{1-\theta_{j1}}{1-\theta_{j0}}\bigg),$ where $N_j(n)$ is the number of units sampled from line $j$,  $S_j(n)$ is the number of success, and $\theta_{j0}$ and $\theta_{j1}$ is the probability of success before and after the change, respectively, $j=1, 2, \cdots, k.$ That is,   $\theta_{j0}$ is the in-control probability and it is assumed known for all $j, j=1, 2, \cdots, k.$ $\theta_{j1}$ is the out-of-control probability and it is normally unknown. Our hypotheses are $H_o: \theta_j=\theta_{j0}$ vs $H_a: \theta_j=\theta^*_{j1},$ where $\theta_{j1}^*$ is a fixed value and it represents the projected out-of-control probability for line $j.$ We use $\theta_{j1}^*$ to calculate $W_j(n)$ to determine the sampling probability $\mbox{\boldmath$\pi$}^*_{n+1}$ based on the Bellman operator $T.$  Our numeric results show the robustness of this method, although the out-of-control probability $\theta_{j1}$ is unknown and the value is being learned from the information in responses collected using the adaptive sampling. 

We use the exact distribution method proposed by Yi (2013) to establish the distribution of $W_j(n)$ to study the statistical power for change detection. For binary outcomes, $W_j(n)$ is a function of  $\{N_1(n), S_1(n), \cdots, N_{k-1}(n), S_{k-1}(n), S_k(n)\}$, $n=1, 2, \cdots,$ where $N_k(n)$ is omitted because $N_k(n)=n-\sum_{j=1}^{k-1} N_j(n).$   This process is a Markov process. The transition of this process is 
$$N_j(n+1)=N_j(n)+\delta_{(n+1)j}, \;\; S_j(n+1)=S_j(n)+\delta_{(n+1)j}Y_{(n+1)j}, j=1, 2, \cdots, k, $$
and the transition probability is $\prod_{j=1}^k \left [\pi^*_{(n+1)j} \theta_j^{Y_{j}}(1-\theta_j)^{1-Y_{(n+1)j}}\right ]^{\delta_{(n+1)j}}, $
 where 
 $\delta_{(n+1)j}$ is sampling result using the sampling probability $\mbox{\boldmath$\pi$}^*_{n+1}$ obtained from the operator $T$ and the initial state is $X_0=(0, 0, \cdots, 0)_{2k-1}.$  Let $X^*=(N^*_1(n+1), S^*_1(n+1), \cdots, N^*_{k-1}(n+1), S^*_{k-1}(n+1)), S^*_k(n+1)$,  $p_{X_0}=1$, and $X_{n+1}=(N_1(n+1), S_1(n+1), \cdots, N_k(n+1), S_k(n+1)).$ The probability to reach $X^*$ is the summation of the probabilities of all the paths of the Markov process to reach it.   That is, 
\begin{eqnarray*}
P(X_{n+1}=X^*) &=& \sum_{\substack{x_{n+1}=X^*, \\ \delta_{ij}, Y_{ij }\\i=1,2,\cdots, n}}\prod_{i=1}^{n+1}\prod_{j=1}^k \left [\pi_{ij} \theta_j^{Y_{ij}}(1-\theta_j)^{1-Y_{ij}}\right ]^{\delta_{ij}} \\
&=& g_{\mbox{\boldmath$\pi$}^*}(X^*, n+1)\prod_{j=1}^k(\theta_j)^{S_j(n+1)}(1-\theta_j)^{N_j(n+1)-S_j(n+1)},
\end{eqnarray*}
where $$g_{\mbox{\boldmath$\pi$}^*}(X^*, n+1)=\sum_{\substack{X_{n+1}=X^*, \\ \delta_{ij}, Y_{ij}\\i=1,2,\cdots, n}} \prod_{i=1}^{n+1}\prod_{j=1}^k \pi_{ij}^{\delta_{ij}}.$$

Therefore, the adaptive sampling strategy is based on  the Markov process $\{N_1(n), S_1(n), \cdots, N_k(n), S_k(n)\}$, $n=1, 2, \cdots.$ The probability for the corresponding value of $W_j(n)$ at $X^*$ is $P(X_{n+1}=X^*)$ , thus the exact distribution of $W_j(n)$ can be established. We use the exact distribution to identify the upper and lower control bounds  to control the probability of false alarm when there are no changes as well as to examine the statistical power to detect a change.

\section{The numeric studies for binary outcomes}

We demonstrate the performance of the adaptive sampling method for binary outcomes using the exact distribution of the Markov process for two lines in a system. The algorithm can be generalized to three or more lines using the method in Yi (2013). Assume that there is a line out of control but we do not know which one is. The adaptive sampling method is to use the incurring information in $W_j(n), j=1, 2,$ to identify which one is more likely to be out of control thus  increasing sampling probability of the line sequentially. Denote the lower control limit and upper control limit as LCB and UCB, respectively. 
Those limits are determined based on the exact distribution of $W_j(n)$ that satisfy the following, assuming $H_o$,
\begin{eqnarray}
& &LCB = Max\{c: P(W_j(n) \leq c) \leq 0.00135\},\\
& & UCB =Min\{c:  P(W_j(n) \geq c) \leq 0.00135 \} .
\end{eqnarray}

We use the same false alarm rate (FAR) $\alpha=0.0027$  as that used in the construction of the Shewhart control charts (Shewhart 1931) for each line to determine LCB and UCB in the numeric studies. Our proposed method can be applied for other values of FAR $\alpha$, where $\alpha/2$ is used to determine LCB and UCB.

Suppose the in-control probability, denoted as $\theta_0$, is the same for both of the two lines. The null hypothesis is $H_o: \theta_{j0}=\theta_0, j=1 ,2. $ Let $L_1$ and $L_2$ be the standardized lower and upper control bounds, respectively, computed using the mean and standard deviation of $W_j(n)$ under $H_o$. The statistical powers are the probability that either lines 1 or 2 cross the bounds when there is a line out of control. That is, the statistical power is computed as 
$P( W_1\leq LCB \;or\; W_1\geq UCB)+P(W_2\leq LCB\; or\; W_2\geq UCB)-P(\;both\; W_1 \; and \; W_2 \mbox{ are outside of the bounds}),$ assuming $H_a.$ 

We consider the scenarios of $\theta_0 = 0.05, 0.1, 0.15$ and $\theta_{j1}^* = 0.1, 0.15, 0.2,$ respectively in the calculation of $W_j(n)$. Suppose that only one line is out of control, say line $1$. We report the expected proportion of sample units $E(N_1/n)$ and the statistical power to detect the change for various scenarios of values of $\theta_{11}.$ Those values are computed using the exact distribution of $W_j(n)$ based the Markov process described in Section 3. 

Suppose the first two units are allocated to the two lines using completely blocking randomization, one on each. The remaining units are allocated using the adaptive sampling based on the operator $T.$ The algorithm to obtain the adaptive sampling probabilities based on the Markov decision process is as follows.

\begin{description}
\item Step 1. The first two units are randomized to the two line, one on each. Produce the state space $\Omega_1=\left\{(1, 1, 1), (1, 1, 0), (1, 0, 1), (1, 0, 0)\right\}.$ For $X=(x_1, x_2, x_3)\in \Omega_1$, the corresponding likelihood $l_X=g_{\mbox{\boldmath$\pi$}}(x,1)\theta_1^{x_2} (1-\theta_1)^{(1-x_2)} \theta_2^{x_3}(1-\theta_2)^{(1-x_3)},$ where $g_{\mbox{\boldmath$\pi$}}(x,1)=1.$

\item Step 2. Let $\Omega_l$ be the sate space at time $l.$ For $X \in \Omega_l, l \geq 1,$ where $X=(x_1(l), x_2(l), x_3(l))$, compute $W_1(l)= x_2(n) log\bigg(\frac{\theta_{11}^*/(1-\theta_{11}^*)}{\theta_{0}/(1-\theta_{0})}\bigg)+x_1(l) log\bigg(\frac{1-\theta_{11}^*}{1-\theta_{0}}\bigg)$, $W_2(l)= x_3(l) log\bigg(\frac{\theta_{11}^*/(1-\theta_{11}^*)}{\theta_{0}/(1-\theta_{0})}\bigg)+(l+1 - x_1(l)) log\bigg(\frac{1-\theta_{11}^*}{1-\theta_{0}}\bigg)$  and the approximate optimal values 

$
\hat{u}_{l}=(l+1)[(1-\gamma)\hat{r}_1+\gamma\hat{r}_2]I_{E}+(l+1)[\gamma\hat{r}_1+(1-\gamma)\hat{r}_2]I_{F}
 +  (l+1)[1/2 \hat{r}_1+1/2\hat{r}_2]I_{G},$
where 
 $E=\{W_1(l) > W_2(l)\}$; $F=\{W_1(l) < W_2(l)\}$; and $G=\{W_1(l)=W_2(l)\}$ and the estimation is for $\theta_j, j=1, 2$ using $X.$

\item Step 3. Look one step further to sample one more unit on line 1 and 2. Calculate $v_j=\hat{r}(j)+ \int \hat{u}_{l}(w+s)\tilde{q} (ds|w, \delta_{.j}=1), j=1, 2,$  and determine 
 \begin{equation*}
\pi_{l+1} = \left\{\begin{array}{ll} 1-\gamma & \mbox{ if } v_1 > v_2,\\
					\gamma & \mbox{ if } v_1 < v_2,\\
					1/2 & \mbox{ if } v_1=v_2 .
					\end{array}\right.
\end{equation*}

\item Step 4. Obtain $\Omega_{l+1}$ and $g_{\mbox{\boldmath$\pi$}}(X, l+1) \mbox{ and }l_X$ for $X\in \Omega_{l+1}$: 
\begin{eqnarray*}
&& \Omega_{l+1}  =  \{X: X=X_{l}+u, X_{l}\in \Omega_{l}, u\in \Lambda\}\\
&& g_{\mbox{\boldmath$\pi$}}(X, l+1)=\sum_{\substack{X_{l}+u=X \\ X_{l} \in \Omega_{l} \\ u\in \Lambda}}g_{\mbox{\boldmath$\pi$}}(X_{l}, l)\pi_{l+1} ^{\delta_{l}}(1-\pi_{l+1}) ^{(1-\delta_{l})}\\
&& l_X=g_{\mbox{\boldmath$\pi$}}(X, l+1) \theta_1^{x_2}(1-\theta_1)^{x_1-x2}\theta_2^{x_3}(1-\theta_2)^{n-x_1-x_3}
\end{eqnarray*}
where $\Lambda=\left\{(1, 1, 0), (1, 0, 0), (0, 0, 1), (0, 0, 0)\right\}$ and $X=(x_1, x_2, x_3).$ Combine the states that have the components to obtain $\Omega_{l+1}$ and combine the corresponding $g_{\mbox{\boldmath$\pi$}}(X, l+1)$ and $l_X.$

\item 5. Go back to Step 2 until all $n$ units are allocated. 
\end{description}

The sampling probability $\pi_l$ in the algorithm depends on the value of $\gamma.$ The magnitude of $\gamma$ represents the trade-off between exploration and exploitation. A large value of $\gamma$ means more exploration for  changes. Our previous research demonstrate that a large value of $\gamma$ provides less skewed allocation asymptotically thus a possible gain in statistical power for hypothesis testing in adaptive clinical trials (Yi and wang 2023).  Let $\gamma=0.25$ in the following. This value provides moderate skewed allocation thus certain exploration on change for each line. 

Table 1 summarizes the numeric results for the adaptive sampling for the in control probability $\theta_0=0.05$ for both of the two lines and  the projected probability of having a change occurring $\theta_{11}^*=0.1$ for line 1, assuming no change in line 2.   It includes the expected mean proportion of units $E(N_1/n)$ allocated to line 1, the statistical power to detect a change, and standardized bounds $L_1$ and $L_2$ for various sample sizes. Table 2 reports statistical power for equal randomization that is computed based on binomial distribution. The equal randomization allocates half of the total units to line 1 and half to line 2. The FAR under the column $\theta_{11}=0.05$ is a single line for Tables 1 and 2, which is the same for both lines under $H_o$. Although statistical powers for Tables 1 and 2 are based on the same formula, the further calculation of $P(\;both\; W_1 \; and \; W_2 \mbox{ are outside of the bounds})$ is different for Tables with adaptive sampling and those with equal allocation. For adaptive sampling, the two samples depend with each other in that one sample is larger so the other is smaller for a total fixed size $n.$ The part that both $W_1$ and $W_2$ are outside of the bounds is the overlap part of the two dependent samples for rejecting $H_o.$ For equal ranodmization, the two samples are independent because half of the total size are randomized to line 1 and half to line 2, and the two lines operate independently.

The results reveal that FAR under the adaptive sampling is more close to the nomination level determined by equations (1) and (2) than the equal allocation. It shows that the bounds $L_1$ and $L_2$ are smaller or larger than those with equal randomization because of the variation introduced by the adaptive sampling, but the adaptive sampling allocate higher proportion of units to line 1, the line having a change occurring, and the statistical power to detect a change with the adaptive sampling is generally higher than that with equal randomization except for size $n=10$ and small difference of out-of-control probability $\theta_{11} $ from 0.05, i.e. $\theta_{11}=0.1, 0.2$. 
As the sample size increases, the adaptive sampling allocates more units to line 1 based on the leaning on which line has a change and the statistical power to detect the change increases. When the out-of-control probability $\theta_{11}$ increases, departing from the projected value of 0.1, the adaptive sampling allocates higher proportion of units to line 1 with increasing statistical power to detect the change. 

Similar results are observed for $\theta_0=0.1, \theta_{11}^*=0.15$ in Tables 3 and 4, and for $\theta_0=0.15, \theta_{11}^*=0.2$ in Tables 5 and 6.

\begin{table}[!ht]
\begin{center}
\begin{tabular}{c|cc|cc|cc|cc|cc}
\hline
& \multicolumn{2}{c|}{\underline{Std bound}}&\multicolumn{2}{c|}{\underline{FAR}} &\multicolumn{6}{c}{\underline{$E(N_1/n)$ and Statistical Power}}\\
 \multirow{2}{*}{$n$}  &\multirow{2}{*} {$L_1$ }& \multirow{2}{*}{$L_2$} & \multicolumn{2}{c|}{\underline{$\theta_{11}=0.05$}}  & \multicolumn{2}{c}{\underline{$\theta_{11}=0.1$}} & \multicolumn{2}{c}{\underline{$\theta_{11}=0.2$}} & \multicolumn{2}{c}{\underline{$\theta_{11}=0.3$}}  \\
 & & & $E(N_1/n)$ & FAR& $E(N_1/n)$ & Power&$E(N_1/n)$ & Power&$E(N_1/n)$ & Power\\ \hline
10&-1.1552&5.4737&0.5&0.00845&0.525&0.00800&0.567&0.05840&0.600&0.17711 \\
20&-1.0862&4.8947&0.5&0.00182&0.542&0.01972&0.602&0.17570&0.648&0.47205\\
30&-1.1379&4.7368&0.5&0.00188&0.552&0.02615&0.630&0.28035&0.673&0.67163\\
40&-1.2413&4.6316&0.5&0.00170&0.559&0.03814&0.647&0.40869&0.689&0.81872\\
50&-1.3793&4.5789&0.5&0.00151&0.567&0.04831&0.660&0.51229&0.700&0.89710\\
60&-1.4138&4.3442&0.5&0.00251&0.574&0.06568&0.670&0.61144&0.708&0.94767\\
100&-1.8966&4.2632&0.5&0.00132&0.598&0.11214&0.697&0.85785&0.724&0.99594\\\hline
\end{tabular} 
\caption{False alarm rate (FAR) for a single line and statistical power to detect a change with the adaptive sampling for  $\theta_{0}=0.05, \theta_{11}^*=0.1$}
\end{center}
\end{table}
 
 \bigskip
 \begin{table}[!ht]
\begin{center}
\begin{tabular}{c|cc|c|c|c|c}
\hline
& \multicolumn{2}{c|}{\underline{Std bound}}& \underline{FAR}& \multicolumn{3}{c}{\underline{Statistical Power}}\\
$n$  &$L_1$ & $L_2$ &  $\theta_{11}=0.05$ & $\theta_{11}=0.1$  & $\theta_{11}=0.2$ & $\theta_{11}=0.3$
 \\ \hline
10&-0.5345&3.6316&0.00158&0.00971&0.05901&0.16405 \\
20&-0.7414&3.6316&0.00103&0.01381&0.12178&0.35106\\
30&-0.4729&3.0976&0.00061&0.01333&0.16475&0.48483\\
40&-1.0512&4.1053&0.00033&0.01159&0.19606&0.58377\\
50&-1.1552&3.4737&0.00121&0.03457&0.38406&0.80675\\
60&-1.2586&3.7895&0.00057&0.02639&0.39338&0.84057\\
100&-1.6552&3.5790&0.00076&0.05858&0.69290&0.98176\\\hline
\end{tabular} 
\caption{FAR for a single line and statistical power to detect a change with equal randomization for  $\theta_{0}=0.05, \theta_{11}^*=0.1$}
\end{center}
\end{table}

\bigskip  
 
\begin{table}[!ht]
\begin{center}
\begin{tabular}{c|cc|cc|cc|cc|cc}
\hline
& \multicolumn{2}{c|}{\underline{Std bound}}&\multicolumn{2}{c|}{\underline{FAR}} &\multicolumn{6}{c}{\underline{$E(N_1/n)$ and Statistical Power}}\\
 \multirow{2}{*}{$n$}  &\multirow{2}{*} {$L_1$ }& \multirow{2}{*}{$L_2$} & \multicolumn{2}{c|}{\underline{$\theta_{11}=0.1$}}  & \multicolumn{2}{c}{\underline{$\theta_{11}=0.15$}} & \multicolumn{2}{c}{\underline{$\theta_{11}=0.2$}} & \multicolumn{2}{c}{\underline{$\theta_{11}=0.3$}}  \\
 & & & $E(N_1/n)$ & Power& $E(N_1/n)$ & Power&$E(N_1/n)$ & Power&$E(N_1/n)$ & Power\\ \hline
10&-1.1379&4.7895&0.5&0.00162&0.523&0.00774&0.543&0.01913&0.600&0.07580\\
20&-1.3793&4.5263&0.5&0.00161&0.534&0.01167&0.566&0.03959&0.618&0.19234\\
30&-1.4828&4.2632&0.5&0.00246&0.543&0.01591&0.582&0.06214&0.642&0.32078\\
40&-1.6897&4.2105&0.5&0.00177&0.551&0.01990&0.596&0.09084&0.659&0.45199\\
50&-1.3793&4.5789&0.5&0.00151&0.567&0.02480&0.607&0.12356&0.672&0.57443\\
60&-2.0345&4.1579&0.5&0.00137&0.564&0.02872&0.617&0.15389&0.682&0.67008\\
100&-2.3103&4.0526&0.5&0.00244&0.585&0.05316&0.647&0.30772&0.706&0.90529\\\hline
\end{tabular} 
\caption{FAR for a single line and statistical power to detect a change  with the adaptive sampling for  $\theta_{0}=0.1, \theta_{11}^*=0.15$}
\end{center}
\end{table}

\bigskip
 \begin{table}[!ht]
\begin{center}
\begin{tabular}{c|cc|c|c|c|c}
\hline
& \multicolumn{2}{c|}{\underline{Std bound}}&\underline{FAR}& \multicolumn{3}{c}{\underline{Statistical Power}}\\
$n$  &$L_1$ & $L_2$ &  $\theta_{11}=0.1$ & $\theta_{11}=0.15$  & $\theta_{11}=0.2$ & $\theta_{11}=0.3$
 \\ \hline
10&-0.7759&3.7368&0.00046&0.00269&0.00718&0.03223\\
20&-1.0862&2.7665&0.00015&0.00153&0.00652&0.04749\\
30&-1.2931&3.8947&0.00031&0.03915&0.01836&0.13141\\
40&-1.5172&3.7368&0.00042&0.00633&0.10325&0.22805\\
50&-1.6897&3.6842&0.00046&0.00843&0.04721&0.32338\\
60&-1.8276&3.6842&0.00045&0.01011&0.06151&0.41146\\
100&-2.3620&3.3157&0.00100&0.03103&0.18687&0.77736\\\hline
\end{tabular} 
\caption{FAR for a single line and statistical power to detect a change with equal randomization for  $\theta_{0}=0.1, \theta_{11}^*=0.15$}
\end{center}
\end{table}

\begin{table}[!ht]
\begin{center}
\begin{tabular}{c|cc|cc|cc|cc|cc}
\hline
& \multicolumn{2}{c|}{\underline{Std bound}}&\multicolumn{2}{c|}{\underline{FAR}} &\multicolumn{6}{c}{\underline{$E(N_1/n)$ and Statistical Power}}\\
 \multirow{2}{*}{$n$}  &\multirow{2}{*} {$L_1$ }& \multirow{2}{*}{$L_2$} & \multicolumn{2}{c|}{\underline{$\theta_{11}=0.15$}}  & \multicolumn{2}{c}{\underline{$\theta_{11}=0.2$}} & \multicolumn{2}{c}{\underline{$\theta_{11}=0.3$}} & \multicolumn{2}{c}{\underline{$\theta_{11}=0.4$}}  \\
 & & & $E(N_1/n)$ & Power& $E(N_1/n)$ & Power&$E(N_1/n)$ & Power&$E(N_1/n)$ & Power\\ \hline
10&-1.4138&4.5789&0.5&0.00132&0.521&0.00554&0.559&0.02632&0.591&0.08354\\
20&-1.5517&4.2632&0.5&0.00228&0.531&0.00884&0.587&0.00610&0.631&0.22221\\
30&-1.8276&4.1579&0.5&0.00140&0.539&0.01040&0.607&0.10637&0.655&0.38301\\
40&-2&4.1053&0.5&0.00174&0.546&0.01323&0.623&0.16306&0.672&0.53956\\
50&-2.2414&4.0526&0.5&0.00184&0.553&0.01662&0.636&0.22643&0.684&0.66989\\
60&-2.2759&4.0526&0.5&0.00241&0.558&0.01974&0.644&0.28859&0.693&0.76727\\
100&-2.3793&3.9211&0.5&0.00265&0.577&0.03613&0.675&0.54854&0.714&0.95628\\\hline
\end{tabular} 
\caption{FAR for a single line and statistical power to detect a change  with the adaptive sampling for  $\theta_{0}=0.15, \theta_{11}^*=0.2$}
\end{center}
\end{table}

\bigskip
 \begin{table}[!ht]
\begin{center}
\begin{tabular}{c|cc|c|c|c|c}
\hline
& \multicolumn{2}{c|}{\underline{Std bound}}& \underline{FAR}&\multicolumn{3}{c}{\underline{Statistical Power}}\\
$n$  &$L_1$ & $L_2$ &  $\theta_{11}=0.15$ & $\theta_{11}=0.2$  & $\theta_{11}=0.3$ & $\theta_{11}=0.4$
 \\ \hline
10&-0.9483&4.1053&0.00076&0.00040&0.00251&0.01032\\
20&-1.3621&4.0263&0.00013&0.00100&0.01073&0.05489\\
30&-1.6552&3.4737&0.00061&0.00485&0.01836&0.13141\\
40&-1.8966&3.1579&0.00133&0.01130&0.11451&0.40519\\
50&-2.1207&3.5263&0.00049&0.00605&0.09825&0.41455\\
60&-2.3276&3.3684&0.00079&0.01027&0.15998&0.56925\\
100&-2.5862&3.3684&0.00096&0.01540&0.31678&0.84406\\\hline
\end{tabular} 
\caption{FRA for a single line and statistical power to detect a change with equal randomization for  $\theta_{0}=0.15, \theta_{11}^*=0.2$}
\end{center}
\end{table}

\bigskip
\section{Conclusion}

We study the change detection problem for multiple lines in a system using adaptive sampling and the sampling process is formatted as a Markov decision process under the average reward criterion. The information in responses are used to sequentially modify the sampling probability while using the Bellman operator to approximate the optimal average awards. The state of the Markov decision process for adaptive sampling is based on the likelihood ratio statistic to detect a change. This method applies to both of continuous and binary responses.

Considering moderate to small sample sizes, we demonstrate the performance of the proposed method for binary responses using the exact distribution method. The algorithm for the optimal adaptive sampling is established and the performance of adaptive sampling is compared with the equal allocation. Our numeric results show that adaptive sampling allocates higher proportion of units to the line that is more likely to have a change and the false alarm rate under the adaptive sampling is more close to the nominal level. The learning on which line has a change depends on sample sizes and the difference between out-of-control and in-control probabilities. When the difference between out-of-control probability and in-control probability increases, higher proportion of units are allocated to the line with a change and the statistical power to detect a change increases. For sample sizes of 20 and above, the statistical power under adaptive sampling is better than that of equal allocation.  

\bigskip

\noindent {\bf Acknowledgement}

Yanqing Yi acknowledge research support from the Natural Sciences and Engineering Research Council (NSERC) of Canada. Both authors thank College of Commerce, National Chengchi University for finance support (114H111-30).


\vspace{0.2in}

\end{document}